\documentclass[preprint,showpacs,preprintnumbers,amsmath,amssymb]{revtex4}

\usepackage{graphicx}
\usepackage{dcolumn}
\usepackage{bm}

\begin{document}

\title{Finite-size effects of a left-handed material slab on the image quality}

\author{Long Chen$^1$, Sailing He$^{1,2,*}$ and Linfang Shen$^1$}

\address{$^1$ Centre for Optical and Electromagnetic Research,
State Key Laboratory of Modern Optical Instrumentation, Zhejiang
University, Hangzhou Yuquan 310027, P. R. China
\\ $^2$ Laboratory of Photonics and Microwave
Engineering, Department of Microelectronics and Information
Technology, Royal Institute of Technology, S-16440 Kista, Sweden}

\begin{abstract}
The characteristics of an imaging system formed by a left-handed
material (LHM) slab of finite length are studied, and the
influence of the finite length of the slab on the image quality is
analyzed. Unusual phenomena such as surface bright spots and
negative energy stream at the image side are observed and
explained as the cavity effects of surface plasmons excited by the
evanescent components of the incident field. For a thin LHM slab,
the cavity effects are found rather sensitive to the length of the
slab, and the bright spots on the bottom surface of the slab may
stretch to the image plane and degrade the image quality.
\\\bf{Keywords:} left-handed material, finite size, surface plasmon, cavity

\end{abstract}

\pacs{78.20.Ci, 42.30.Wb, 73.20.Mf}
\maketitle

Recently, a new type of composite material (also called
left-handed material (LHM) since the electric field, the magnetic
field and the wave vector of an electromagnetic plane wave
propagating in such a material obey the left-hand rule) which
exhibits simultaneously negative effective permittivity and
permeability over a certain frequency band has attracted a great
attention \cite{Veselago,Shelby,Pendry,PRLPapers}. The
extraordinary electromagnetic properties of LHM include reverse
Doppler shift, backward-directed Cherenkov radiation cone and
inverse Snell effect, which were first analyzed theoretically by
Veselago \cite{Veselago}.

It has been noticed recently that an infinitely-extended LHM slab
can focus not only the propagating waves from the object, but also
the evanescent waves corresponding to the sub-wavelength structure
of the object. Therefore, theoretically such a LHM slab can
reconstruct the original object and achieve a perfect resolution
under an ideal case (i.e., the so-called perfect lens)
\cite{Pendry}. While the inherent material loss will greatly
suppress the amplification of evanescent waves and make the
perfect image impossible \cite{Garcia}, subwavelength imaging is
still achievable for a thin LHM slab \cite{SmithAPL}. However, in
a realistic imaging system the length of the LHM slab must also be
finite, particularly for applications such as micro-detectors and
micro-imaging devices where the device size is required to be as
small as possible. In this letter, we study the effects of the
finite size of the LHM slab to the imaging quality. A
finite-difference time-domain (FDTD) method \cite{Taflove} is used
in the numerical simulation.

The two-dimensional (2D) imaging system we consider here is a
planar LHM slab (surrounded by vacuum) with a finite length of $L$
and a thickness of $d$. The slab is located in the region of
$(-L/2<x<L/2, 0<z<d)$, and a point source located at ($x=0, z=-u$)
is used to generate the object for the imaging system (see the
inset of Fig. 1(a)). Here we consider only the E-polarization
where $\bf{E}$ is directed in the $y$ direction. For matched
material parameters and $u<d$, the field will be focused at $z=u$
inside the slab and $z=2d-u$ outside the slab \cite{Pendry}. Here
for simplicity we set $u=0.5d$ in all our imaging simulation. To
avoid the field singularity of the point source, the $object$
plane is selected to be slightly ($0.05\lambda$) below the point
source, and the $image$ plane is shifted correspondingly. The
computational domain is bounded by perfect-matched layers (PMLs)
and a FDTD method of scattered-field/total-field version is
adopted \cite{Taflove}. To avoid the divergence (occurred when the
permittivity and permeability are negative) as the time marches in
the FDTD simulation, the following Drude's dispersion model
\cite{Cummer} for the permittivity and permeability of the LHM
slab is used,
\begin{eqnarray}
\varepsilon \left( \omega \right) =\varepsilon _0(1-\frac{\omega
_{pe}^2}{\omega^2}), \quad \mu \left( \omega \right) =\mu
_0(1-\frac{\omega _{pm}^2}{\omega^2})
\end{eqnarray}
The permittivity and permeability will take negative values for
frequencies below $\omega _{pe}$ and $\omega _{pm}$, respectively.
Here we assume $\omega _{pe}=\omega _{pm}$ and the material
parameters are matched (i.e., $\varepsilon \left( \omega_0 \right)
/\varepsilon _0=\mu \left(\omega_0 \right) /\mu _0=-1$; as assumed
for a perfect lens) at frequency $\omega _0 =\omega_{pe}/ \sqrt{2}
$. To minimize the frequency extension, the time-dependence of the
point source is set as $exp(i\omega_0t)f(t)$, where $f(t)$ is a
step function that reaches $1$ smoothly in a time duration of
$30T_0$ (here the period $T_0={2\pi}/{\omega_0}$). The grid size
of the discretization is $0.01\lambda$. After enough time steps,
the field evolution becomes stable and the stable field is taken
as the field at frequency $\omega _0$ (the accuracy has been
verified by taking the Fourier transform of the time sequence of
the field to extract the field at frequency $\omega _0$ and thus
the dispersion effect is negligible for our monochromatic
incidence).

{\bf Unusual Phenomena:} Interesting phenomena can be observed
clearly from Fig. 1(a) for the distributions of the normalized
field intensity and $z$ component of the energy stream ($S_z$) on
the image plane. Here we choose $L=8\lambda$ and $d=0.2\lambda$.
Unlike the ideal imaging for an infinitely-extended LHM slab, next
to the central peak (the desired image) the image for a LHM slab
of finite length has additional peaks with considerable magnitudes
(even exceeding the magnitude of the central peak in some cases).
More surprisingly, near the central peak the energy stream $S_z$
has large negative values (reaching about $-20\%$ after
normalization for this example). The negative $S_z$ on the image
plane seems counter-intuitive since there is no scatterer below
the slab and the normal energy stream there $should$ simply flow
downward (i.e., positive $S_z$) from the slab. The corresponding
2D distribution of the field intensity is shown in Fig. 1(b).
Clearly one sees many bright spots (nearly equi-distanced)
distributed along each surface of the LHM slab. Although the field
decreases exponentially away from these surface spots, they still
stretch to the image plane due to their large magnitudes and
consequently damage the image. The two additional peaks in Fig.
1(a) on the image plane are the extension of the two brightest
spots centered on the bottom surface of the slab (see Fig. 1(b)).
We have increased $d$ (up to $3\lambda$) and varied $L$ (from
$1.5\lambda$ to $9\lambda$) and these unusual phenomena are still
observed.

{\bf Cavity Effects:} In the imaging system of infinitely extended
LHM slab with matched material parameters, incident evanescent
wave will experience amplification inside the slab and thus have
strong intensity around the exit (i.e., bottom) interface of the
slab (see e.g. \cite{Pendry}). This wave can be conceived as a
localized field around this single interface and is exponentially
decreasing both into the slab and below the slab. Therefore, the
simple term ``surface plasmon'' (commonly used for a
surface-localized wave in metal-related electromagnetism) is
``borrowed'' to describe this evanescent wave with large field
amplitude near the interface. For a point source located above the
slab, the field intensity on the upper or bottom surface has a
simple profile with only one central peak. Neither the additional
surface bright spots nor negative energy stream can be observed
for such an infinitely-extended LHM slab. However, when the slab
has a finite length, each excited surface plasmon can be roughly
conceived as travelling along the bottom surface and encountering
a side-end (i.e., the left or right boundary) of the slab. The
main part of the energy should be reflected, with the remaining
part running across the corner or coupling to radiation. The
original and the multiply reflected waves are superimposed to form
a standing wave profile. Therefore, the side-ends act as
reflecting walls and the finite slab behaves like an
one-dimensional cavity for the surface plasmons. This lateral
cavity effect leads to the distribution of nearly equi-distanced
bright spots (due to the nodal structure of the standing wave)
along the exit interface of the LHM slab. Meanwhile, the energy
stream $S_z$ along the surface of the slab has an oscillating
behavior (with a period similar to that of the surface bright
spots) and takes negative values at some positions, as shown in
Fig. 1(c). Due to the continuity of the normal stream $S_z$ on the
bottom surface, these negative streams extend to the image plane
and cause negative values at the image side. Note that the meaning
of ``surface plasmon'' used here differs from that of ``surface
polariton'' (or ``slab plasmon polariton''), which usually refers
to the coupled surface waves on both interfaces, i.e., symmetrical
or anti-symmetrical eigen modes of the slab. Such eigen modes do
not exist for a LHM slab with an infinite length and matched
material parameters (see e.g. \cite{Ruppin,SmithAPL}).

Here we show that these bright spots and negative energy streams
result from the resonance of surface plasmons, which are excited
only by the evanescent components of the incident wave. Filters
for the spatial spectrum are employed to extract the propagating
or evanescent components of the incident field. The incident
fields are first Fourier transformed with respect to $x$. A window
function is applied to these spectra and the modified spectra are
then transformed back to the physical space through the inverse
Fourier transform. Therefore, a low-pass window function with the
upper truncation $k_x=k_0$ will extract the propagating components
of the incident field, while a high-pass window function with the
lower truncation $k_x=k_0$ will extract the evanescent components.
For graphic clarity we consider here a slab with $L=2\lambda$ and
$d=2u = 0.2\lambda$. Note that the length of this slab is not too
short since $k_x$ has a large value (corresponding to a small
period of oscillation in $x$ direction. See e.g., there are about
nine nodes over the slab length of $2\lambda$ in Fig. 2(a) below).
The 2D field intensity distributions contributed by the full
spectrum, the propagating parts and the evanescent parts of the
incident field are shown in Figs. 2(a-c), respectively. The field
intensity contributed by the propagating components (Fig. 2(b)) is
rather simple, with no reflection at the upper and bottom
interfaces (as expected), since under present matched material
parameters the propagating components should experience simple
negative refraction at the upper and bottom interfaces, and the
ends of the finite slab causes only some negligible scattering.
The small side-lops are due to the limited wavenumbers for the
propagating parts ($k_x<k_0$). However, the field intensity
contributed by the evanescent components (Fig. 2(c)) has a similar
distribution of surface bright spots (only slight difference in
their relative magnitudes) as compared with Fig. 2(a) for the case
of full spectrum. We have also compared the energy streams along
the image plane for these cases. The stream $S_z$ has a simple
profile and is non-negative everywhere when only the propagating
components are included, while oscillating behavior and negative
values are observed when the evanescent components are included.

We can also numerically simulate the resonant behavior of a single
evanescent wave for our imaging system. The 2D distributions of
the field intensity and energy stream for the case of
$L=2\lambda$, $d= 0.2\lambda$ and $k_x=2k_0$ are shown in Figs.
2(d) and 2(e), respectively. Equi-distanced bright spots along the
surface with a period of ${\pi}/{k_x}=0.25\lambda$ are clearly
seen in Fig. 2(d), while in Fig. 2(e) the energy stream $S_z$
exhibits an oscillating property with alternating positive and
negative values along the $x$ direction. As $L$ varies, we do
observe that the resonance strength (measured by the maximal field
intensity $|E|^2$ along the bottom surface) oscillates with $L$
with a period around $0.25\lambda$ and has multiple peak values,
resembling the resonant behavior of a simple one-dimensional
resonator.

In addition to the lateral resonance caused by the reflection at
the side-ends as analyzed above, the bright spots along the upper
surface (see e.g. Fig. 2(d)) suggest a coupling mechanism between
the upper and bottom surfaces through the slab ends, and can be
explained here by a dynamic procedure. For a LHM slab of infinite
length with matched material parameters, the amplification of a
single evanescent wave should produce strong field intensity only
around the bottom surface. However, for a LHM slab of finite
length, along the bottom surface the amplified evanescent wave
diffracts near the side-ends and the corners and couples some part
of energy back into the slab. Like an additional evanescent wave
incident from the other side of the slab, the diffracted wave will
also experience an amplification inside the slab and lead to an
enhanced field along the upper surface. Similar diffractions
happen recursively around both surfaces until a stable field
distribution with bright spots on both surfaces is established,
introducing a cavity effect also along the $z$ direction. The
field evolution in our FDTD simulation verifies this dynamic
procedure, and under some cases (e.g. $L=2.2\lambda$,
$k_x=2.8k_0$) the field strength along the upper interface is much
stronger than that along the bottom interface.

Here we present some numerical analysis of the two-dimensional
cavity effects (i.e. the dependence of the resonant strength on
the wave-vector $k_x$, the length $L$ and the thickness $d$). Fig.
3(a) shows the resonant strength of different $k_x$ for
$d=0.2\lambda$ and $L=2.0\lambda$ (diamonds), $2.2\lambda$
(circles) and $2.5\lambda$ (squares), respectively. Here to
balance the amplification inside the slab for a fair comparison
between different $k_x$, we keep the source-slab distance $u=d$.
As mentioned above, the resonant strength depends greatly on $L$,
and it also has a selective effect on $k_x$ with multiple resonant
peaks (marked by arrow, other peaks beyond the range of this
figure are observed and relatively have much weaker amplitude).
Additionally, these cases share the same peak positions (see e.g.
$k_x \approx 2.25k_0$), which suggests that the resonant $k_x$
(which excites the strongest resonance along the slab surfaces)
shall be related to the coupling between the two slab surfaces
(i.e., the other dimension $d$ of the proposed cavity), and the
slab length $L$ mainly influences the amplitude. As shown in Fig.
3(b) for $L=2.0\lambda$, the resonant peak positions shift
monotonously from $k_x \approx 2.6k_0$ to $2.25k_o$ and $1.65k_0$
for $d$ from $0.15\lambda$ (diamonds) to $0.2\lambda$ (circles)
and $0.3\lambda$ (squares), and their corresponding peak values
decrease for about one order for $d$ from $0.15\lambda$ to
$0.3\lambda$.

An incident field generated by a point source contains many
evanescent waves. Due to the sensitivity of the resonant strength
of each evanescent component to length $L$, the overall field for
the case of a point source is complex and depends greatly on $L$.
A simple comparison for the field intensity profiles along the
bottom surface and the corresponding images are shown in Fig. 4(a)
and (b) for two slightly different values of $L$ (i.e.,
$L=2\lambda$ and $2.2\lambda$), respectively. Here the thickness
is still kept as $d =0.2\lambda$. Much brighter surface spots
exist for $L=2\lambda$, and along the image plane the two unwanted
large peaks around $x= \pm 0.23 \lambda$ are the extension of the
two distinct surface spots centered on the bottom surface of the
slab. Contrarily, the case of $L=2.2\lambda$ gives a flattened
central spot and weaker field in the side-lop region, and thus the
image quality is better as compared with that for $L=2\lambda$. As
we increase $L$ further, the field intensity profiles are found to
exhibit a quasi-periodic behavior. For $L=2.4\lambda$, the image
quality becomes poor again with two unwanted peaks greatly
exceeding the central peak. The additional surface spots still
exist and degrade the image quality for $L$ up to $9\lambda$, and
the image quality varies between the best cases of clear image
£¨e.g. $L=8.2\lambda$£©with the full width at half maximum (FWHM)
about $0.24\lambda$ (the FWHM of the object is about
$0.22\lambda$) and the worst cases (e.g. $L=2.4\lambda$) that the
expected point image is greatly exceeded by the additional
resonant peaks (by more than 130\%). However, the distance between
these resonant peaks has greatly decreased from about
$0.23\lambda$ to $0.11\lambda$ when we increase $L$ from
$2\lambda$ to $9\lambda$. For a sufficiently long slab, the
resonant spectrum shall tend to become flat and these lateral
nodal structures are expected to partially cancel each other
except for the central peak (i.e., the desired image). Thus, the
influence of the cavity effects on the image decreases gradually
and the field intensity distribution gradually transits to the
simple shape for a LHM slab of infinite length. For a thicker
slab, the resonance becomes weaker and more flat (see Fig. 3(b)),
and its influence on the image quality decreases because of its
evanescent characteristics along the $z$ direction and the larger
distance between the image plane and the bottom surface of the
slab. It is also less sensitive to the length of the slab. In our
simulation for $d=\lambda$, along the image plane the intensity of
these additional peaks are always less than $30\%$ of that of the
central peak (the desired image) for $L>7\lambda$, and the
negative streams are less than $-2\%$, producing a clear point
image with FWHM about $0.3\lambda$. Similarly, as we have
verified, the material loss of the LHM slab (if included) also
damps the excitation of surface plasmons and their cavity effects.

In conclusion, the imaging system formed by a LHM slab of finite
length has been analyzed through a FDTD method. The cavity effects
of surface plasmons excited by the evanescent components of the
incident field have been studied and used to explain the observed
bright spots along the surfaces of the LHM slab and
counter-intuitive negative values of the normal component of the
energy stream at the image side of the slab. The extended bright
spots on the bottom surface of the LHM slab may stretch to the
image plane and degrade the image quality. It has been shown that
both the length and thickness of the LHM slab greatly influences
the cavity effects of surface plasmons and consequently the image
quality.

{\bf Acknowledgments:} The partial support of National Natural
Science Foundation of China (under a key project grant; grant
number 90101024) is gratefully acknowledged. The authors are also
grateful to the referees for their valuable comments.

$*$ Electronic Address: sailing@tet.kth.se

\newpage

\begin{figure}
\includegraphics[width=7cm]{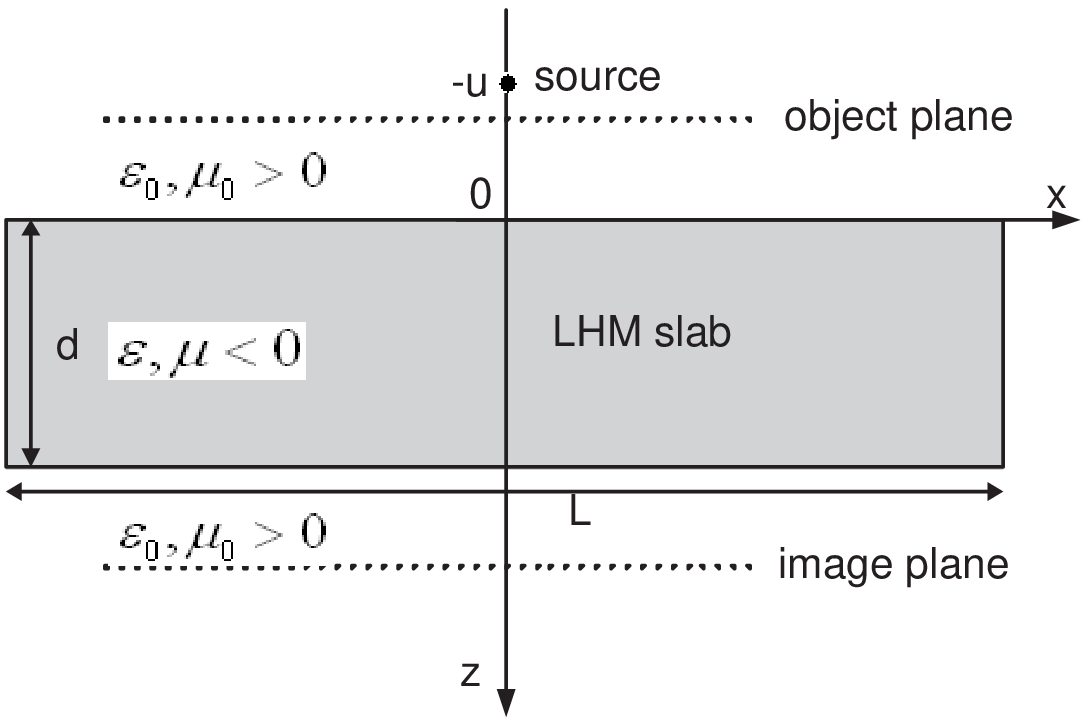}
\includegraphics[width=7cm]{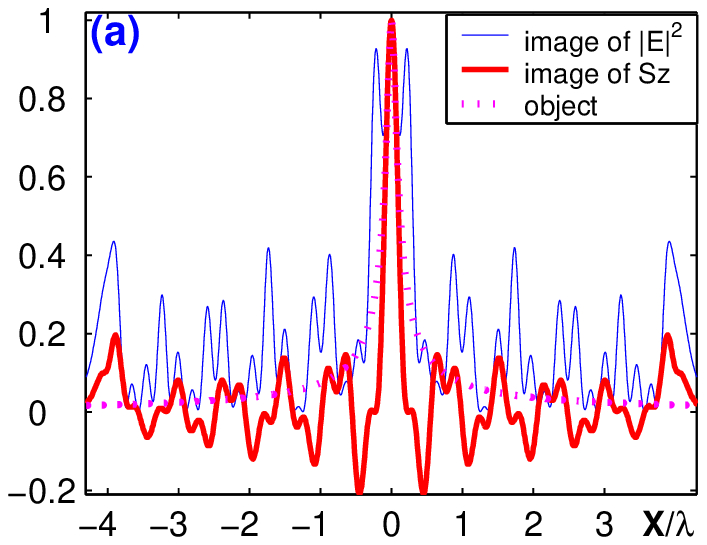}
\includegraphics[width=7cm]{fig1bc}
\caption{(a) Normalized distribution of field intensity (thin
line) and the energy stream $S_z$ (thick line) along the image
plane for $L=8\lambda$ and $d=0.2\lambda$; here the normalized
field intensity of the object is plotted (dotted line) for
comparison. Inset: the geometry of the imaging system formed by a
LHM slab of finite length. The corresponding 2D distribution of
(b) $|E|^2$ and (c) $S_z$. Here the LHM slab and the image plane
are marked with a rectangle and a dashed line, respectively.}
\end{figure}

\begin{figure}
\includegraphics[width=7cm]{fig2abc}
\end{figure}
\begin{figure}
\includegraphics[width=7cm]{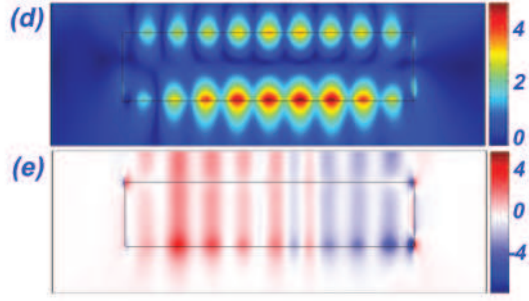}
\caption{(a-c) 2D distribution of the field intensity after
applying a filter to the spatial spectrum of the incident field
generated by a point source: (a) with the full spectrum (b) with
only the propagating parts (c) with only the evanescent parts of
the spatial spectrum. (d) and (e) give the 2D distributions of the
field intensity and $S_z$, respectively, when the resonance is
excited by a single evanescent wave with $k_x=2k_0$. Here
$L=2\lambda$ and $d=2u = 0.2\lambda$.}
\end{figure}

\begin{figure}
\includegraphics[width=7cm]{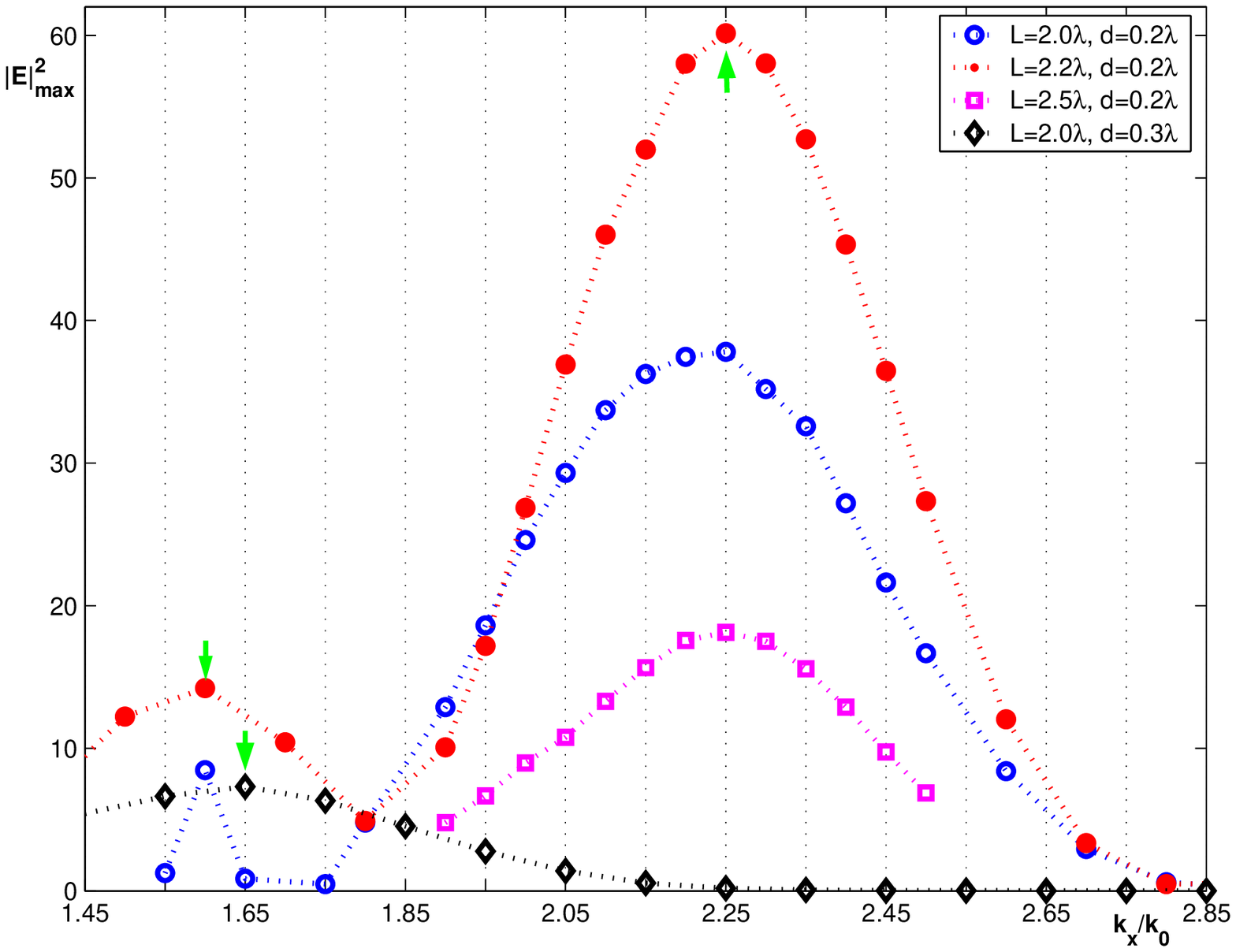}
\includegraphics[width=7cm]{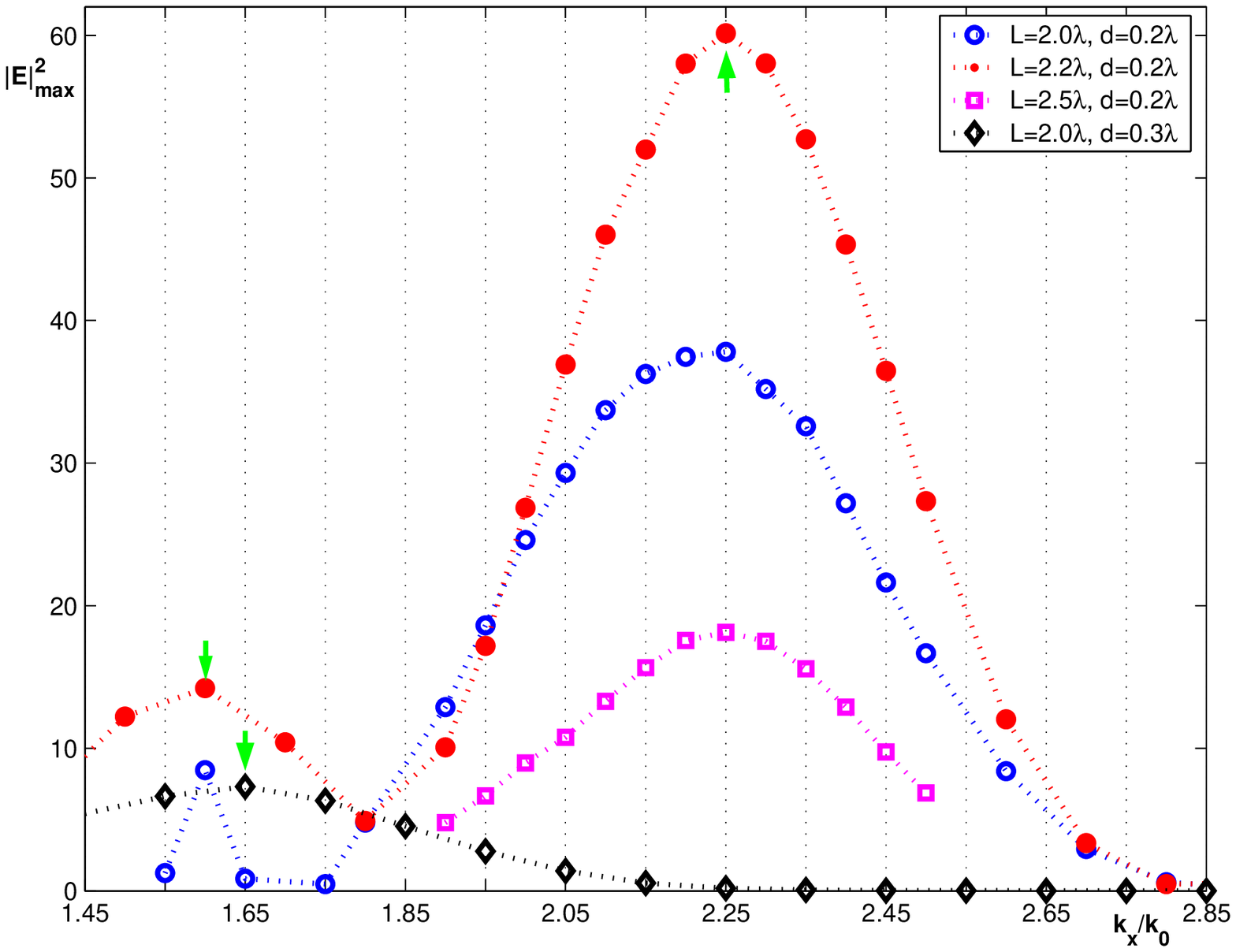}
\caption{Dependence of the resonant strength on $k_x$: (a) for
$d=0.2\lambda$ and $L=2.0\lambda$ (diamonds), $2.2\lambda$
(circles) and $2.5\lambda$ (squares), respectively; (b) for
$L=2.0\lambda$ and $d=0.15\lambda$ (diamonds), $0.2\lambda$
(circles) and $0.3\lambda$ (squares), respectively.}
\end{figure}

\begin{figure}
\includegraphics[width=7cm]{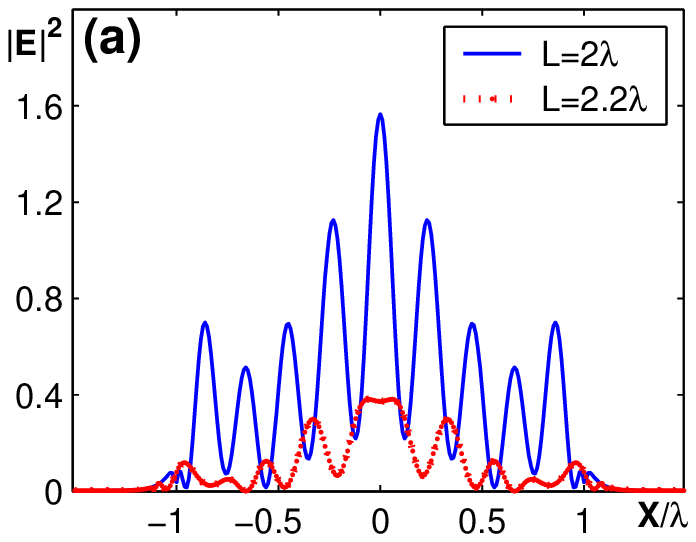}
\includegraphics[width=7cm]{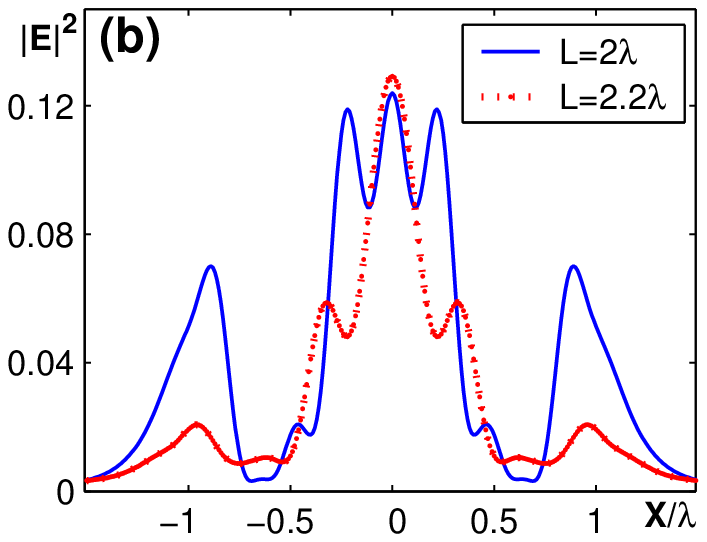}
\caption{Field intensity profiles (a) along the bottom surface of
the LHM slab and (b) along the image plane, when the incident
field is generated by a point source. Here $d=2u = 0.2\lambda$.}
\end{figure}

\end{document}